\documentstyle[preprint,prd,aps,tighten,epsf]{revtex}
\begin{document}
\draft
\newcommand{\nl}{\nonumber \\}
\newcommand{\bea}{\begin{eqnarray}}
\newcommand{\eea}{\end{eqnarray}}
\newcommand{\bi}{\bibitem}
\newcommand{\be}{\begin{equation}}
\newcommand{\ee}{\end{equation}}
\newcommand{\bt}{\begin{table}}
\newcommand{\et}{\end{table}}
\newcommand{\btab}{\begin{tabular}}
\newcommand{\etab}{\end{tabular}}
\def\eg{{\it e.g.}}
\def\ra{\rightarrow}
\def\dek#1{\times10^{#1}}
\def\Vus{\left|V_{us}\right|}
\def\vp{{\bf p}}
\def\piee{K^+\ra\pi^+e^+e^-}
\def\pimm{K^+\ra\pi^+\mu^+\mu^-}
\def\pill{K^+\ra\pi^+\ell^+\ell^-}
\def\die{e^+e^-}
\def\dim{\mu^+\mu^-}
\def\tautripi{\tau^-\ra\pi^-\pi^-\pi^+\nu_\tau}
\def\taua1{\tau^-\ra a_1^-\nu_\tau}
\def\kmu{K^+\ra\mu^+\nu_\mu}
\def\ratio{\dim/\die}
\def\rms{m_\rho^2(t)}
\def\mrs{m_\rho^2}
\def\mg{m_\rho\Gamma(t)}

\def \zpc#1#2#3{Z.~Phys.~C {\bf#1}, #2 (19#3)}
\def \plb#1#2#3{Phys.~Lett.~B {\bf#1}, #2 (19#3)}
\def \prl#1#2#3{Phys.~Rev.~Lett.~{\bf #1}, #2 (19#3)}
\def \pr#1#2#3{Phys.~Rep.~{\bf #1}, #2 (19#3)}
\def \prd#1#2#3{Phys.~Rev.~D~{\bf#1}, #2 (19#3)}
\def \npb#1#2#3{Nucl.~Phys.~B {\bf#1}, #2 (19#3)}
\def \rmp#1#2#3{Rev.~Mod.~Phys.~{\bf#1}, #2 (19#3)}
\def \ea{{\it et al.}}
\def \ibid#1#2#3{{\it ibid.} {\bf#1}, #2 (19#3)}
\def \sjnp#1#2#3#4{Yad. Fiz. {\bf#1}, #2 (19#3) [Sov. J. Nucl. Phys.
{\bf#1}, #4 (19#3)]}
\def \pan#1#2#3#4{Yad. Fiz. {\bf#1}, #2 (19#3) [Physics of Atomic Nuclei
{\bf#1}, #4 (19#3)]}
\def \epjc#1#2#3{Eur. Phys. J. C {\bf#1}, #2 (19#3)}
\def \appb#1#2#3{Acta Phys. Pol. B {\bf#1}, #2 (19#3)}

\title{\bf 
Running mass of the $\rho^0$ meson's implication for the dilepton mass 
spectrum and the $\ratio$ branching ratio in the $\pill$ decays}

\author{Peter Lichard\footnote{On leave of absence from Department of 
Theoretical Physics, Comenius University,
842-15 Bratislava, Slovak Republic.}}
\address{
Department of Physics and Astronomy, University of Pittsburgh,
Pittsburgh, Pennsylvania 15260 \\
and Institute of Physics, Silesian University, 746-01 Opava, Czech Republic
}
\maketitle
\begin{abstract}
We make an attempt to resolve the discrepancy of the observed $\die$ mass 
spectrum in the $\piee$ decay with that predicted by meson dominance.
To this end we investigate the properties of the $\rho^0$ propagator. We use 
dispersion relations to evaluate the running mass $\rms$ of the $\rho^0$ 
resonance without adjustable parameters. To improve the convergence of the
dispersion integral, the momentum dependence of strong vertices is taken 
from the flux-tube-breaking model of Kokoski and Isgur. The obtained behavior 
of $\rms$ at small momentum squared $t$ makes the $\piee$ form factor rise 
faster with
increasing $t$ than in the original meson dominance calculation and more 
in agreement with the published data. As a consequence, the meson dominance 
prediction of the $\ratio$ branching 
ratio changes slightly, from 0.224 to 0.236. 
We do not see any possibility to accommodate into the meson dominance 
approach an even steeper $\die$ spectrum, indicated by the preliminary 
data of the E865 collaboration at BNL AGS.
\end{abstract}

\pacs{PACS number(s): 12.15. y, 12.15.Ji, 11.55.Fv, 13.20.Eb}

\narrowtext

The decays $\pill$ ($\ell=e\ ,\mu$) have been the subject of intensive 
theoretical studies since the late fifties (see 
\cite{gaillard,nanopoulos,vain76} and references therein). The picture of 
later theoretical development can be gained by inspecting 
\cite{berg,ecker87,fajfer96,ambrosio} and papers cited there. The decay 
$\pill$ was experimentally observed in  1975 in its $\die$ mode \cite{bloch} 
and in 1997 in the $\dim$ mode \cite{adler}. Other experiments include
a more precise measurement of the $\die$ mode by the BNL-E777 collaboration
\cite{alliegro}, unpublished $\die$ data of the BNL-E851 collaboration 
\cite{e851}, and the current BNL-E865 experiment, capable of measuring
both modes \cite{hong} with high precision and statistics.

Today, it is customary to interpret the experimental results in the 
framework of the chiral perturbation theory
\cite{ecker87,donoghue,ambrosio}. Unfortunately, this theoretical framework 
contains free parameters, one in the $p^4$ order \cite{ecker87}, 
two in the $p^6$ order \cite{ambrosio}. This, on one hand, diminishes 
the predictive power of the theory but, on the other hand, gives more 
room to experimentalists when trying to fit theoretical formulas.

\begin{figure}
\begin{center}
\leavevmode
\setlength \epsfysize{12cm}
\setlength \epsfxsize{15cm}
\epsffile{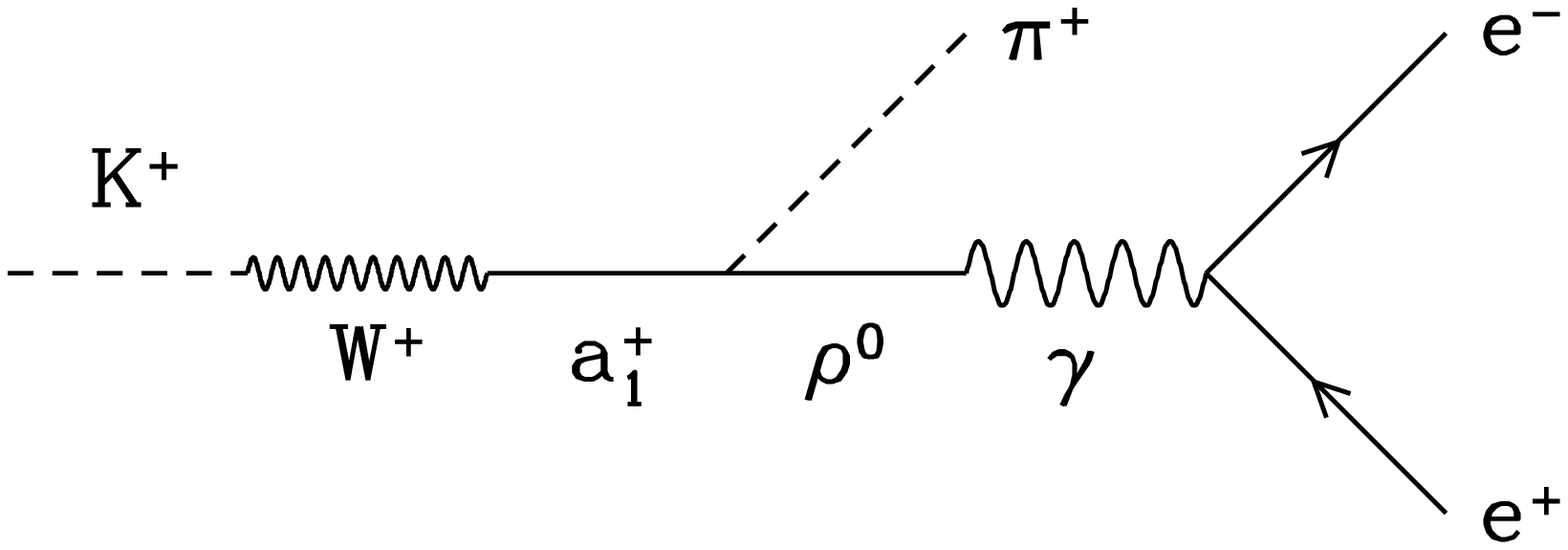}
\end{center}
\caption{Matrix element of the decay $K^+\ra\pi^+e^+e^-$ in
the meson dominance approach.}
\label{figure1}
\end{figure}

On the contrary, as we have shown recently \cite{md}, the meson 
dominance offers a parameter-free description of the $\pill$ decays.
The relevant Feynman diagram is shown in Fig.~\ref{figure1}. The 
corresponding formula for the differential decay rate in dilepton mass $M$
has the form generally expected for the one-photon approximation, namely
\be
\label{kpiee}
\frac{d\Gamma(\pill)}{dM}=
C\ M\lambda^{3/2}(m_{K^+}^2,m_{\pi^+}^2,M^2)\
\sqrt{1-\frac{4m_\ell^2}{M^2}}\left(1+\frac{2m_\ell^2}{M^2}\right)
\left|F(M^2)\right|^2
\ee
with $\lambda(x,y,z)=x^2+y^2+z^2-2xy-2xz-2yz$ and the form factor 
given by
\be
\label{naiveff}
F(t)=\frac{m_\rho^2}{m_\rho^2-t}\ .
\ee
The normalization constant $C$ is not given by first principles, 
but can be determined using the data other than those on the $\pill$ 
decays themselves. Concretely, by the experimental information about the 
$\tautripi$ and $\kmu$ decay. In Ref.~\cite{md}, we used the decay rate
of the $\taua1$ decay, the $a_1(1270)$ decay width, and the $\kmu$
branching fraction. In this way we obtained $B(\piee)\approx 3.1\dek{-7}$, 
not in contradiction with experiment $(2.74\pm0.23)\dek{-7}$
\cite{pdg}. The approximative character
of our result was caused by the badly known $a_1$ decay width.

Formula (\ref{kpiee}) makes a definite prediction for the $\ratio$ branching
ratio even if $C$ is badly known. The number is 0.224  with an error which 
is negligible under the circumstances because the ratio is a function of the 
masses of participating particles only. We use it and the experimental
$\die$ branching fraction to predict
\be
\label{origpred}
B(\pimm)=(6.2\pm0.5)\dek{-8},
\ee
in agreement with the later measurement \cite{adler} of 
$(5.0\pm1.0)\dek{-8}$.\footnote{We follow the convention of \cite{pdg},
where the statistical, systematic and theoretical errors given in
\cite{adler} are summed quadratically.} 

The $t$-dependence of the form factor (\ref{naiveff}) can be, for the
purpose of comparing with data, characterized by the slope variable
\be
\label{lambda}
\lambda(t)=m_{\pi^+}^2\frac{dF(t)}{dt}\ ,
\ee
which is equal to 0.033 at $t=0$ and reaches 0.053 at the upper
kinematic boundary $t\approx 0.125$~GeV$^2$. In Ref.~\cite{alliegro}
the data were fit by a linear approximation to the form factor
[$\lambda\equiv\lambda(0)$]
\be
\label{linear}
F(t)=1+\lambda\frac{t}{m_{\pi^+}^2}
\ee
with $\lambda=0.105\pm0.035\ ({\rm stat.})\pm0.015\ ({\rm syst.})$.

This result became a little surprising after the experimental value 
of $B(\pimm)$ was published \cite{adler}. In fact,
if one assumes the $\mu/e$ universality and validity of (\ref{linear}),
the values of $B(\piee)$, $B(\pimm)$, and $\lambda$ must match together. 
And they do not match very well. Even for $\lambda=0.055$ (mean value
minus both errors) the ``predicted" value of the $\dim/\die$
branching ratio is equal to 0.235, which should be compared to the
experimental $0.18\pm0.04$ (using \cite{pdg}, errors summed quadratically).
The disagreement rises with $\lambda$. The preliminary data of the E865 
experiment \cite{hong} indicates that the fault is not on the
side of $\lambda$.

We therefore take for granted that the experimental value of 
$\lambda$ indicates, despite its large errors, that the meson dominance 
form factor (\ref{naiveff}) is too flat. In the following, we will try
to find the possible origin of this discrepancy and a way to improve
the situation without introducing unnatural assumptions and free parameters.
 
When writing (\ref{kpiee}) with form factor given by (\ref{naiveff}), the 
essential assumption was that the $\rho^0$ propagator in Fig.~\ref{figure1} 
can be written in a free-vector-particle form 
\be
\label{freepropagator}
-iG^{\mu\nu}_0(q)=\frac{-g^{\mu\nu}+q^\mu q^\nu/m_\rho^2}{t-m_\rho^2+
i\epsilon}\ ,
\ee
where $m_\rho$ is the mass of the $\rho^0$ resonance, as it is seen in the 
hadronic production experiments. The general expression for the 
interacting-vector-resonance propagator is a little more complicated. It
sounds (see, \eg, \cite{isgurtau})
\be
\label{vresprop}
-iG^{\mu\nu}(q)=\frac{-g^{\mu\nu}+\omega(t)q^\mu q^\nu/t}
{t-m_\rho^2(t)+im_\rho\Gamma(t)}\ ,
\ee
where $\Gamma(t)$ is the total width of the $\rho$-resonance with
off-shell mass $\sqrt t$, normalized at $t=m_\rho^2$ to the nominal 
width $\Gamma_\rho$. Furthermore, $\rms$ is the running mass squared and
$\omega(s)$ is a complex function which reflects the properties of the 
one-particle-irreducible bubble.

      The propagator (\ref{freepropagator}), which is usually used 
in meson dominance calculations, differs from (\ref{vresprop}) in
three respects:
\begin{enumerate}
\item A simplified structure of the $q^\mu q^\nu$ term. This is not
important, because this term does not contribute anyhow due to
the transverse $\rho\pi\pi$ vertex.
\item The absence of a finite imaginary part, which is justified, since most 
of our $t$-region lies below the $\pi\pi$ threshold. In a small window
between the latter and the end of the $t$-interval it is negligible.
Nevertheless, we will include it in what follows.
\item The only real difference is in replacing the running mass 
$m_\rho(t)$ with the nominal mass $m_\rho$, what is generally
allowed only in a close vicinity of the resonance point. 
\end{enumerate}

We will concentrate our effort on the last issue and study the
consequences of replacing nominal mass of the $\rho^0$ resonance by its 
running mass in the denominator in Eq.~(\ref{naiveff}). To be more 
concrete, we will write the modified form factor in the form 
\be
\label{modifiedff}
F(t)=\frac{m_\rho^2(0)}{m_\rho^2(t)-t-im_\rho\Gamma(t)}\ . 
\ee

It follows from the causality of the propagator (\ref{vresprop}) that
above the $\pi\pi$ threshold $t_0$, the $\rms$ and $\mg$ are boundary values
of the real and imaginary parts, respectively, of a function analytic in the 
cut $t$-plane. We can therefore write a once subtracted dispersion relation
\cite{isgurtau}
\be
\label{dispersion}
\rms=\mrs-\frac{t-\mrs}{\pi}\ {\cal P}\!\int_{t_0}^\infty\frac{m_\rho
\Gamma(t^\prime)}{(t^\prime-t)(t^\prime-\mrs)}\ dt^\prime\ ,
\ee
where symbol ${\cal P}$ denotes the principal value. To proceed further,
we must find all important contributions to the variable
width $\Gamma(t)$. Without any doubt, we start with $\rho^0\ra\pi^+\pi^-$.
Other candidates are, ordered according to rising thresholds: $\rho^0\ra
\eta\pi^+\pi^-$, $\rho^0\ra\omega\pi^0$, and $\rho^0\ra K^+K^-\&   
K^0\bar{K}^0$.  The relative importance of those channels can
further be assessed by comparing the abundance of their isotopic companions
in the $\tau^-$ decays. This suggests that of those three, the 
$\omega\pi^0$ final state
will be most important, while the $\eta\pi^+\pi^-$ one least important.
The results of actual calculations confirm this estimate. Furthermore,
the inspection of the $\tau-$ decay fractions shows that there
is no other important hadronic channel with quantum numbers of the
$\rho^0$ meson. In addition, we assume that possible channels with the 
thresholds above the $\tau^-$ mass may be neglected. The results obtained 
below seem to validate this assumption. 

Now we going to describe our calculation in more detail. Let's start 
with the most important contribution to $\Gamma(t)$, which is the 
$\rho^0\ra\pi^+\pi^-$ decay. We write the $\rho\pi\pi$ vertex in the form
\be
\label{rhopipi}
V^\mu=f_{\rho\pi\pi}({p^*}^2)\left(p_{\pi^+}^\mu-p_{\pi^-}^\mu\right)\ ,
\ee
where $p^*$ is the pion momentum in the $\rho$ rest frame.
Instead of the usual coupling constant we have introduced the strong form 
factor. Its momentum dependence was taken from the flux-tube-breaking
model of Kokoski and Isgur \cite{kokoski}. We thus write
\be
\label{psqdep}
f_{\rho\pi\pi}({p^*}^2)=g_{\rho\pi\pi}
\exp\left\{-\frac{{p^*}^2}{12\beta^2}\right\}
\ee
with $\beta=0.4$. We must confess that our original motivation for borrowing
(\ref{psqdep}) from \cite{kokoski} was technical: we just wanted
to ensure a good convergence of the dispersion integral. But it appeared
later that the very reasonable result for $\rms$, which we will present 
below, could not be achieved without assuming (\ref{psqdep}) or with a very 
different value of parameter $\beta$. Our opinion is now that the 
flux-tube-breaking model Ansatz (\ref{psqdep}) reflects correctly the real 
dynamics of the $\rho\pi\pi$ vertex. We will use the same parametrization
for all strong form factors.

Using (\ref{rhopipi}) and (\ref{psqdep}) we easily arrive at the formula
\be
\label{gammapipi}
\Gamma_{\rho^0\ra\pi^+\pi^-}(t)=\frac{g_{\rho\pi\pi}^2}{6\pi}
\frac{{p^*}^3}{t}
\exp\left\{-\frac{{p^*}^2}{6\beta^2}\right\}\ ,
\ee
where $p^*=\sqrt{t/4-m_{\pi}^2}$. The coupling constant was determined
from the condition 
\be
\label{gsqcond}
\Gamma_{\rho^0\ra\pi^+\pi^-}(m_\rho^2)=\Gamma_\rho=
(150.7\pm1.1)\ {\rm MeV}
\ee
with the result $g^2_{\rho\pi\pi}=41.7\pm0.3$. Formula (\ref{gammapipi})
can be used, with obvious modifications also for $\rho^0\ra K\bar{K}$.
Here, the coupling constant can be determined from the $\tau^-\ra
K^-K^0\nu_\tau$ branching fraction. We refer the reader to Ref.~\cite{md}.
Taking into account the modifications connected with present usage
of the momentum dependent strong form factors and assuming that 
$\rho^-$ and $\rho^0$ decay to their corresponding $K\bar{K}$ systems
with the same rate, we get
$g^2_{\rho^0 K^+K^-}+g^2_{\rho^0 K^0\bar{K}^0}=28.2\pm5.1$.

The $\rho\omega\pi$ vertex is taken in the form\footnote{The coupling
constants here were made dimensionless, contrary to \cite{md}, by 
introducing the $\rho$ mass in the denominator.}
\be
\label{rhoompi}
V^{\mu\nu}=\frac{f_{\rho\omega\pi}({p^*}^2)}{m_\rho}
\ \epsilon^{\mu\alpha\nu\beta}p_{\rho,\alpha}p_{\omega,\beta} 
\ee
with the same momentum dependence of the strong form factor as in
(\ref{psqdep}). The coupling constants can be determined from decay rate
$\Gamma(\omega\ra\pi^0\gamma)=(7.2\pm0.4)\dek{-4}$~GeV assuming usual 
vector-meson-dominance form of coupling between $\rho^0$ and $\gamma$.
The result is $g^2_{\rho\omega\pi}=155\pm8$. The contribution to
$\Gamma(t)$ is given by the formula
\be
\label{gammarhoompi}
\Gamma_{\rho^0\ra\omega\pi^0}(t)=\frac{g^2_{\rho\omega\pi}}
{12\pi}\ \frac{{p^*}^3}{t}\
\exp\left\{-\frac{{p^*}^2}{6\beta^2}\right\}\ ,
\ee
with ${p^*}^2=\lambda(t,m_\omega^2,m_{\pi^0}^2)/(4t)$.

The last contribution to $\Gamma(t)$ we consider is the decay
$\rho^0\ra\eta\pi^+\pi^-$. We will considered as a two-step process:
$\rho^0\ra\eta\rho^0$ followed by the decay $\rho^0\ra\pi^+\pi^-$.
The mass squared of the parent $\rho^0$ is $t$, that of the daughter 
$\rho^0$ is $s<t$. Thanks to the daugher decay matrix element being
transverse, the decay rate of the whole process factorizes into
two parts \cite{aps}. The first of them is given by formula
(\ref{gammarhoompi}) with obvious modifications; the
second one contains a Breit-Wigner term with  decay rate of
$\rho^0\ra\pi^+\pi^-$. The only new element is the $\rho\eta\rho$
coupling constant, which is determined from $\Gamma(\rho^0\ra
\eta\gamma)=(3.6\pm1.3)\dek{-5}$~GeV as $g^2_{\rho\eta\rho}=55\pm21$. 
The contribution to $\Gamma(t)$ is given by
\be
\label{gametapipi}
\Gamma_{\rho^0\ra\eta\pi^+\pi^-}(t) = \frac{g^2_{\rho\eta\rho}
g^2_{\rho\pi\pi}}{36\pi^3\ t}\ \int_{2m_{\pi^+}}^{\sqrt{t}-m_\eta}
\frac{(p_\pi^*p_\eta^*)^3}{(s-m^2_\rho)^2+m_\rho^2\Gamma^2(s)}
\exp\left\{-\frac{{p^*_\pi}^2+{p^*_\eta}^2}{6\beta^2}\right\}\ d\sqrt{s}\ ,
\ee
where
\bea
p_\eta^* &=& \frac{\lambda^{1/2}(t,s,m_\eta^2)}{2\sqrt{t}} \ ,\\
p_\pi^*  &=& \sqrt{\frac{s}{4}-m_{\pi^+}^2} \ .
\eea
It seems to be a sort of conundrum that the right hand side
contains the same quantity, the contribution to which we aim to 
determine, namely $\Gamma(s)$. Under different circumstances we would
be forced to repeat whole procedure several times in search for a
selfconsistent solution. Fortunately, here it shows that the result
depends only little on the form of $\Gamma(s)$. We compared the
case of fixed width $\Gamma_\rho$ with the case of $\Gamma_{\rho\pi\pi}(s)$ 
and found only tiny differences. We picked the result of the latter choice.

\begin{figure}
\begin{center}
\leavevmode
\setlength \epsfysize{15cm}
\setlength \epsfxsize{15cm}
\epsffile{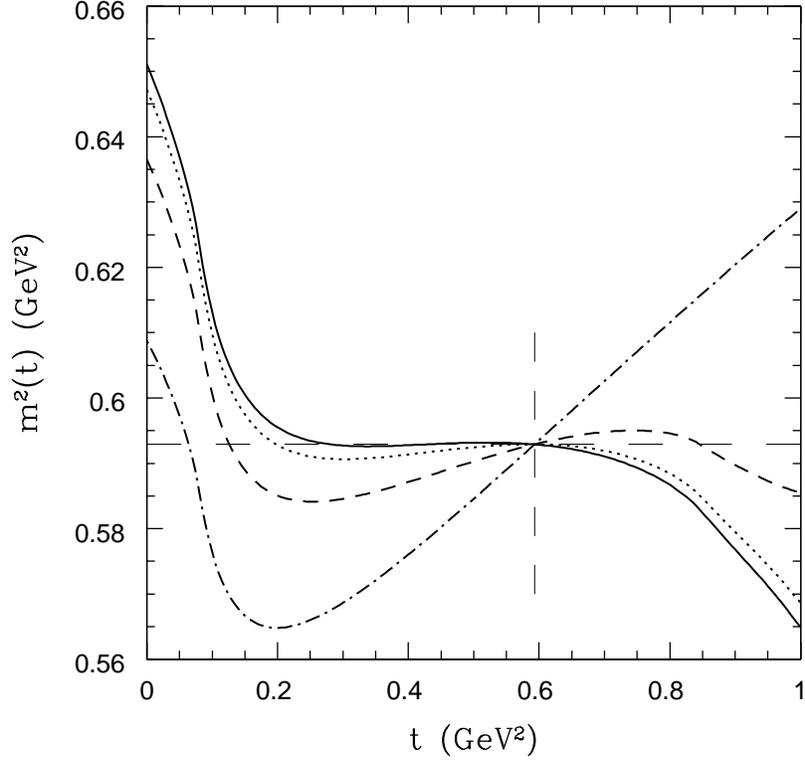}
\end{center}
\caption{Running mass squared of the $\rho^0$ meson for different
inputs to the dispersion relation: (i) $\rho^0\ra\pi^+\pi^-$ only
(dash-dotted curve); (ii) $\rho^0\ra\omega\pi^0$ added (dashed);
(iii) also $\rho^0\ra K\bar{K}$ added (dotted curve); (iv) the final curve
(solid) after the $\rho^0\ra\eta\pi^+\pi^-$ contribution has been added.
}
\label{figure2}
\end{figure}

Now we have collected all pieces and can add them to form the total
$\Gamma(t)$ and evaluate the dispersion integral. To be sure that we
have things under control, we proceeded in a less straightforward
way. We first took the basic ($\rho^0\pi^+\pi^-$) contribution alone and
determined $\rms$. Then we did the same thing for the basic contribution
combined with three other contributions taken individually and
compared the changes against the basic contribution alone. In this
way we determined the following sequence of contributions (most
important first): $\omega\pi^0$, $K\bar{K}$, $\eta\pi^+\pi^-$.
Then we started again and added the contributions cummulatively, in
the order just shown. The resulting $\rms$'s are depicted in 
Fig.~\ref{figure2}. We can see that the procedure of adding contributions 
converges to a very reasonable result: a wide plateau, the derivative at 
$t=m_\rho^2$ almost vanishing. Our input parameters, coupling constants,
have relatively large errors. So it would be possible to vary them
within limits in effort to find even better solution (characterized
by the vanishing derivative at $t=m_\rho^2$. Another possibility would
be to fiddle with $\omega$ a little around the breaking-flux-tube model
\cite{kokoski} preferred solution $\beta=0.4$. We made only one try
in that direction. We found that the derivative vanished if the 
coupling constants squared of all three additional contributions were
diminished by 8\%. But the behaviour of the running mass squared in
the region which interests us most [$0<t<(m_{K^+}-m_{\pi^+})^2$] did
not change by that move at all. We therefore believe that our determination
of $\rms$ at low $t$ is stable and trustable.

Before we draw conclusions about the $\pill$ form factor, we must
mention one correction we should make in order 
to be consistent with the formalism we used in our dispersion relation
evaluation of the $\rho^0$ running mass. We should include the same momentum
dependence of strong form factors also into our basic diagram, 
Fig.~\ref{figure1}. Here, it applies to the $a_1\rho\pi$ vertex and leads
to the following modification of the form factor (\ref{modifiedff}):
\be
\label{ffwithvertex}
F(t)=
\frac{m_\rho^2(0)}{m_\rho^2(t)-t-im_\rho\Gamma(t)}\
\exp{\left\{\frac{t(2m_{K^+}^2+2m_{\pi^+}^2-t)}{48m_{K^+}^2\omega^2}\right\}}
\ee

\begin{figure}
\begin{center}
\leavevmode
\setlength \epsfysize{15cm}
\setlength \epsfxsize{15cm}
\epsffile{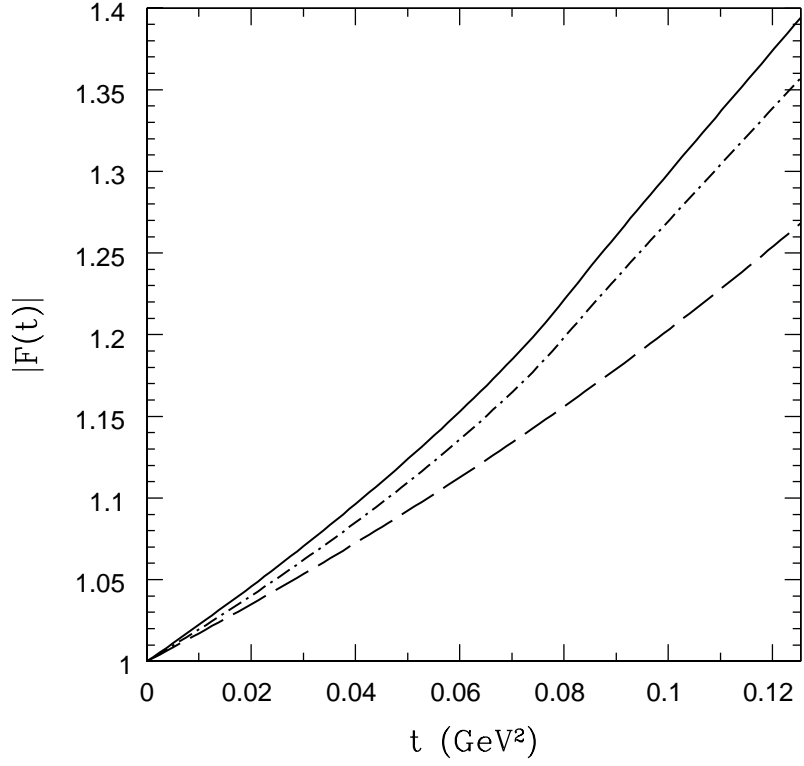}
\end{center}
\caption{$\piee$ form factor as a function of $t$: (i)
constant $\rho^0$ mass, Eq.~(\protect\ref{naiveff}) (dashed); (ii)
running $\rho^0$ mass, Eq.~(\protect\ref{modifiedff}) (dash-dotted); 
(iii) running $\rho^0$ mass and the $a_1\rho\pi$ vertex correction,
Eq.~(\protect\ref{ffwithvertex}) (solid).}
\label{figure3}
\end{figure}

Anyhow, to see the effect of the running mass alone,  in
Fig.~\ref{figure3} we present three curves: the old meson dominance
form factor calculated from Eq.~(\ref{naiveff}) (dashed curve); the form 
factor coming from the running mass with the vertex correction ignored,
Eq.~(\ref{modifiedff}) (dash-dotted curve); and the form factor reflecting
both effects, Eq.~(\ref{ffwithvertex}) (solid curve). The latter is
what we consider the final product of our study.

\begin{figure}
\begin{center}
\leavevmode
\setlength \epsfysize{15cm}
\setlength \epsfxsize{15cm}
\epsffile{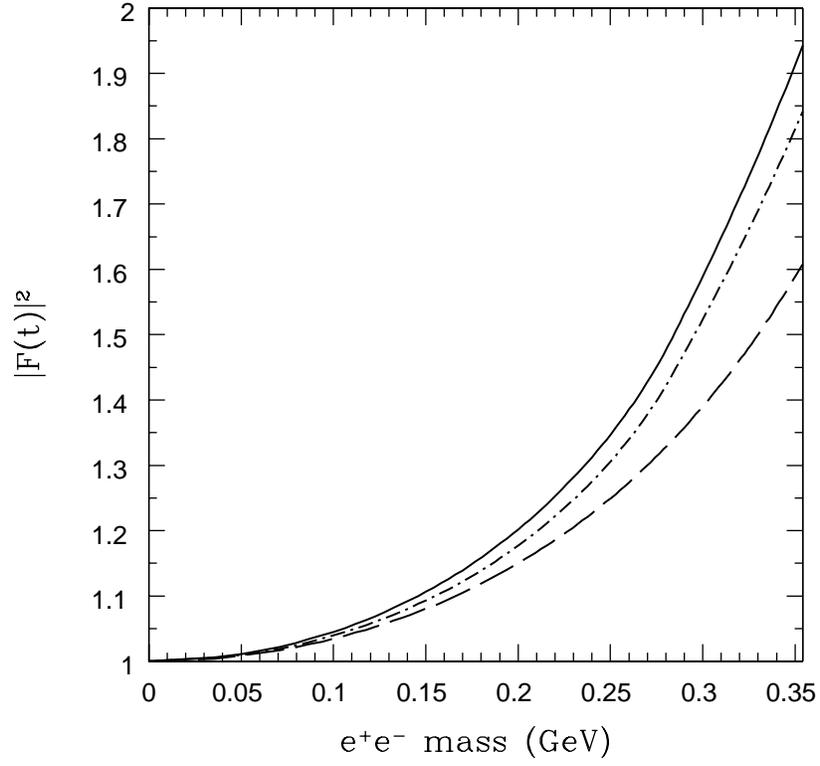}
\end{center}
\caption{Same as Fig.~\protect\ref{figure3}, but the $\piee$ form factor 
squared as a function of the dielectron mass.} 
\label{figure4}
\end{figure}

Fig.~\ref{figure4} brings the same information but in a form which
is better suited for comparison with the experimental mass spectra.
It shows the dependence of the form factor squared on the dilepton mass.

The form factor calculated from (\ref{ffwithvertex}) has a much steeper
$t$ dependence than the original form factor (\ref{naiveff}). It is
characterized by $\lambda=0.043$ at $t=0$ and $\lambda=0.073$ at 
the largest $t$. The $\die/\dim$ branching ratio calculated using 
(\ref{kpiee}) with (\ref{ffwithvertex}) is 0.236. Using the experimental
branching fraction of the $\die$ mode \cite{pdg} we get a new prediction
for the $\dim$ mode
\be
B(\pimm)=(6.5\pm0.6)\dek{-8},
\ee
which differs only little from the original one (\ref{origpred}).
The ``effective $\lambda$" of our form factor, defined as the value
of $\lambda$ in linear parametrization (\ref{linear}) that leads to
the same $\dim/\die$ branching ratio, is 0.057 (in original version of the
meson dominance calculation \cite{md} it was 0.039). From the above
we conclude that the meson dominance model with our new form factor 
is consistent with the shape of the $\die$ mass distribution
as measured in experiment \cite{alliegro}.

A much different story is the comparison with the preliminary data 
\cite{hong} of the E865 experiment at the Brookhaven National Laboratory
Alternating Gradient Synchrotron. Their 10,000 $\piee$ events yielded
a preliminary result of the form factor parameter $\lambda=0.20\pm0.02$.
If this value is confirmed, the meson dominance model of the
$\pill$ decays will be ruled out, despite its success with a parameter-free
calculation of the branching fractions.
 
Let us conclude with a general comment. 
It is a little unfortunate that the role of the $\pill$ decays is sometimes
shrunk to a testing ground of the chiral perturbation theory and
other clues are not followed. Here I mean mainly the importance of these
decays in studying the behavior of the electromagnetic form factor
induced by the $\rho^0$ resonance at small $t$. In my opinion, there
are only two kinds of processes that are able to perform this task. Besides
the $\pill$ decays these are the $\omega\ra\pi^0\ell^+\ell^-$ Dalitz 
decays.\footnote{The Dalitz decays of $\phi$, which can, in principle, 
serve for the same purposes, have not been observed yet.}
Concerning the latter, the only $\die$ experiment performed
\cite{omegadalee} had low statistics and was unable to bring the mass
spectrum. The $\dim$ experiment \cite{omegadalmm} showed that the dimuon mass
spectrum disagreed with the vector meson dominance hypothesis. The parallel
with $\pill$ is interesting. But the kaon decays we consider here are unique 
in populating mainly the region below the $\pi\pi$ threshold, whereas
the dilepton mass spectrum of the $\omega$ Dalitz decays spans to much higher
values. The $\pill$ decays can serve as a unique magnifying glass
for studying behavior of the $\rho$-induced electromagnetic form factor 
at small $t$.

\acknowledgements
I am indebted to Julia Thompson, Naipor Cheung, Dave Kraus, Hong Ma, 
and Pavel Reh\'{a}k for 
discussions and to Julia Thompson and Alex Sher for taking on themselves 
a part of my experimenter's responsibilities while I was working 
as a theoretician. This work was supported by the U.S. Department of Energy 
under contract No. 
DOE/DE-FG02-91ER-40646 and by the Grant Agency of the Czech Republic under
contract No. 202/98/0095.

\end{document}